\documentclass[12pt]{article}
\usepackage{latexsym}
\usepackage{epsfig}

\usepackage{graphicx}
\usepackage{epstopdf}

\usepackage{epsfig,amssymb,amsmath,amscd}

\newcommand{\bq}{\begin{equation}}
\newcommand{\eq}{\end{equation}}
\newcommand{\bqa}{\begin{eqnarray}}
\newcommand{\eqa}{\end{eqnarray}}
\newcommand{\ben}{\begin{enumerate}}
\newcommand{\een}{\end{enumerate}}
\newcommand{\bc}{\begin{center}}
\newcommand{\ec}{\end{center}}
\newcommand{\bqb}{\begin{eqnarray*}}
\newcommand{\eqb}{\end{eqnarray*}}

\def\pr#1#2#3{Phys. Rev. ${\bf{#1}}$, #2 (#3)}

\def\pl#1#2#3{Phys. Lett. ${\bf{#1}}$, #2 (#3)}

\def\np#1#2#3{Nucl. Phys. ${\bf{#1}}$, #2 (#3)}

\def\epj#1#2#3{Eur. Phys. J. ${\bf{#1}}$, #2 (#3)}

\def\jmp#1#2#3{J. Mod. Phys. ${\bf{#1}}$, #2 (#3)}

\begin{document}
\pagenumbering{arabic}
\thispagestyle{empty}
\def\thefootnote{\fnsymbol{footnote}}
\setcounter{footnote}{1}

\begin{flushright}
June 13, 2017\\
 \end{flushright}

\begin{center}
{\Large {\bf CSM analyses of $t\bar t H, t\bar t Z, t\bar b W^-$
production in gluon-gluon and photon-photon collisions}}.\\
 \vspace{1cm}
{\large F.M. Renard}\\
\vspace{0.2cm}
Laboratoire Univers et Particules de Montpellier,
UMR 5299\\
Universit\'{e} Montpellier II, Place Eug\`{e}ne Bataillon CC072\\
 F-34095 Montpellier Cedex 5, France.\\
\end{center}

\vspace*{1.cm}
\begin{center}
{\bf Abstract}
\end{center}

We extend our previous analysis of $t\bar t H, t\bar t Z, t\bar b W$
production in $e^+e^-$ collisions within the CSM (Composite Standard Model)
concept of Higgs boson and top quark compositeness to the case of gluon-gluon
and photon-photon collisions. We show that remarkable differences may indeed
appear between the effects generated by CSM conserving and by CSM violating
sets of form factors.

\vspace{0.5cm}

\def\thefootnote{\arabic{footnote}}
\setcounter{footnote}{0}
\clearpage

\section{INTRODUCTION}

In view of the search of signals of Higgs and/or top quark compositeness
we have recently considered the $e^+e^- \to t\bar t H, t\bar t Z, t\bar b W^-$
processes \cite{CSMee}.
We made this analysis within a concept that we called Composite Standard Model
(CSM) which assumes that compositeness preserves the main SM properties at
low energy, for example the gauge structure and the Goldstone equivalence.
We do not base our analyses on a specific model, we only assume that such a model
may exist using the various proposals of Higgs boson and the top quark compositeness
(based on substructures or on additional sets of states), see
ref.\cite{comp, Hcomp2, Hcomp3, partialcomp, Hcomp4}.
In this spirit, compositeness may reveal itself through the presence of form factors which
are close to one at low energy and do not modify the SM properties.
But as the energy increases the essential CSM property is the preservation
of the typical SM combinations ensuring a good high energy behaviour
 \cite{WLZL}. This leads
to specific relations (that we called CSM constraints) between several
form factors in the Higgs and top quark sectors.\\
For the $e^+e^- \to t\bar t H, t\bar t Z, t\bar b W^-$ processes \cite{CSMee}
we gave illustrations showing remarkable differences between the effects of CSM
conserving and of CSM violating form factors. The most spectacular case
concerns the W production process in which the above mentioned combinations
(ensuring the cancellations of unitarity violating contributions)
are the most striking.\\
In this paper we extend this type of analysis to $t\bar t H, t\bar t Z, t\bar b W^-$
production in gluon-gluon and photon-photon collisions.
We make illustrations for each process. We show how these processes
may give complementary results to those of the $e^+e^-$ set in particular
for distinguishing effects from form factors satisfying or violating CSM constraints.\\
In these illustrations we only show the ratios of new cross sections over SM ones.
We do not discuss the quantitative observability of the modifications because
we do not have yet a precise model to test.\\

Contents: In Sect.2 we present the SM diagrammatic contents of each process.
In Sect.3 we recall the choices of CSM conserving and CSM violating form factors
already used in the previous $e^+e^-$ case. In Sect.4 we show and discuss
the corresponding illustrations for gluon-gluon and photon-photon collisions.
Conclusions are given in Sect.5.\\

\section{TREATMENT OF THE $gg, \gamma\gamma \to  t\bar t H, t\bar t Z, t\bar b W^-$ PROCESSES}

The corresponding diagrams are drawn in Fig.1 for gluon-gluon collisions
and in Fig.2 for photon-photon.\\

\underline{$ t\bar t H$ production}\\

This proceeds through two types of diagrams. The first one corresponds to $t\bar t$
production by top echange and $H$ emission from the final lines or from the intermediate
top. Symmetrization with respect to the two initial beams is applied.\\
The second type only occurs in the gluon-gluon case where the $t\bar t$ pair is
produced by an intermediate gluon and $H$ emission appears from the $t$ or
from the $\bar t$ line.\\

The couplings that can be affected by Higgs and top compositeness are
$gtt$, $\gamma tt$ and $Htt$.\\
A specific effective form factor may be attributed to each of them as discussed
in the next section, with or without CSM constraint.\\
In the illustrations we will introduce
$F_{tR}(s)$, $F_{tL}(s)$ (for simplicity the same for gluon and photon) and $F_{Htt}(s)$
form factors.\\

\underline{$t\bar t Z$ production}\\

The diagrams are exactly of the same type as in the above $ t\bar t H$ case.
The $Htt$ coupling is now replaced by the left and right $Ztt$ couplings
with corresponding $F^Z_{tR}(s)$, $F^Z_{tL}(s)$ form factors (again for simplicity
chosen similar to $gtt$ and $\gamma tt$ ones).\\

\underline{$t\bar b W^-$ production}\\

There is now an essential difference between gluon-gluon and
photon-photon collisions. In the gluon-gluon case the set of
diagrams is again the same as in the $ t\bar t H$, $ t\bar t Z$ processes,
with now only the pure left-handed $Wtb$ coupling.\\
In the photon-photon case one has 4 additionnal diagrams (with the necessary
symmetrization) involving W exchange and the 3-leg or 4-leg photon-W couplings.
In the CSM spirit with the $W_L-G_L$ equivalence, Higgs compositeness
suggests the presence of form factors in these couplings.\\

\section{Test form factors}

We will make illustrations in the same way as in ref.\cite{CSMee}
by affecting to each top and Higgs coupling form factors
satisfying or violating the CSM constraints.\\
We use the same "test" expression
\bq
F(s)={(m_Z+m_H)^2+M^2\over s+M^2}~~\label{FF}
\eq
\noindent
with the new physics scale $M$ taken for example with the value of 4 TeV
chosen for making the illustrations clearer.\\

CSM conserving cases will be denoted as "CSM..." and CSM violating cases
as "CSMv...". In addition, for $W^-_L$ production, we will consider
special CSM conserving cases, denoted "CSMG...", which assume that
(apart from $m^2/s$ suppressed contributions) the amplitudes remain
equivalent to those for $G^-$
production even with the effects of CSM conserving form factors.\\

Precisely in Fig.3-9 we find:\\

--- CSMtLR and CSMGtLR refering to the presence of $t_{L}$ and $t_{R}$ left and right
top form factors put in any gauge (gluon, photon, Z and W) coupling,
and simultaneously Higgs and Goldstone -top coupling form factor
satisfying the CSM relation recalled in eq.(1) of \cite{CSMee}.\\

--- CSMtR and CSMGtR for pure $t_R$ compositeness, $F_{tR}(s)\equiv F(s)$
while $F_{tL}(s)=1$,
and  an effective top mass $m_t(s)=m_tF(s)$ controlling all Higgs sector couplings
and the top kinematical contributions, see \cite{trcomp}.\\

--- CSMvt, violating CSM constraints because of different form factors; 
$F_{tR}(s)$ with $M=8$ TeV differs from
$F_{tL}(s)$ with $M=12$ TeV (also present with the $Wtb$ coupling)
and from
$F_{VW^+_LW^-_L}(s)=F_{Htt}(s)=F_{Gtt}(s)\equiv F(s)$ with $M=4$ TeV,
the top mass being fixed at its on-shell value.\\

--- CSMvH for the case of no top form factor and only one  $F(s)$ form factor affecting the
$VW^+_LW^-_L$ vertices, the top mass being also fixed at its on-shell value.\\

We recall that the aim of these choices is only to illustrate the sensitivity
of the observables to compositeness effects. The compositeness structure may
be much richer (excited states, resonances,..., see for example \cite{res}),
which would require a more involved effective adequate form factor.\\

\section{DISCUSSION OF RESULTS}

In fig.3-9 illustrations are given for the ratios of cross sections
(in the center of the 3-body final state) with form factor effects over pure
SM cross sections.\\

\underline{$t\bar tH$}\\

The energy dependence shapes are similar in gg (Fig.3a) and in $\gamma\gamma$ collisions
(Fig.3b) (of course when assuming that the $gtt$ and the $\gamma tt$ form
factors are similar).\\

As compared to the $e^+e^-$ case, \cite{CSMee} one observes a stronger decrease
with energy due to the presence of additionnal top form factors in gluon or photon couplings.\\

We can then also compare the different CSM conserving or CSM violating cases.
The ordering of the size of the effects, CSMvH/CSMtR/CSMvt/CSMtLR from the less affected case
to the strongest one, corresponds to the simultaneous presence of different
form factors.\\

\underline{$t\bar tZ$}\\

We now illustrate separately the $Z_L$ (Fig.4a,b), the $Z_T$ (Fig.5a,b)
and the unpolarized $Z$ case (Fig.6a,b), with (a) for gluon-gluon and 
(b) for $\gamma \gamma$ process. \\

The various shapes are due to different left, right top couplings form factors and to
the possible use of an effective top mass.\\
Note that in the gluon-gluon process CSMvH produces no effect as no
3-boson coupling is involved in this process.\\

For $Z_L$ the ordering is now
CSMvH/CSMvt/CSMtR/CSMtLR
because of the possible presence of a specific  $Z_L=G$ form factor
and the additional decrease due to an effective top mass in the
CSMtR case.\\

For $Z_T$ without these properties the ordering is 
CSMvH/CSMtR/CSMvt/CSMtLR. \\

The unpolarized case corresponds clearly to the addition of the  $Z_L$
and $Z_T$ contributions with their respective weights.\\

\underline{$t\bar b W^-$}\\

Illustrations are given in (Fig.7a,b) for $W^-_L$ ,
(Fig.8a,b) for $W^-_T$ and (Fig.9a,b) for unpolarized $W^-$. \\

As in other  $W^-$ production processes, because of the presence 
of addtional diagrams the complicated procedure of cancellation 
for the $W^-_L$ component plays an important role in the resulting 
effects of the presence of form factors.
When form factors introduce temporary differences between growing
contributions this generates an increase of the $W^-_L$ amplitude.
For example the fact that the top quark may be composite and
have a form factor whereas the bottom quark remains elementary
without form factor creates an important lack of cancellation.
The ordering is then specific of each process
because on the one hand form factors lead to a decrease with the energy
whereas on the other hand the lack of cancellation leads to an increase.
This is why we consider for this process also the Goldstone equivalent
CSMGtLR and CSMGtR sets.\\
The CSM assumption of $(Z_L,W^-_L)-(G^{0},G^{-})$ equivalence ensures a good
high energy behaviour (even with new effects) such that the presence
of form factors produces immediately a decreasing effect.
We will see below the difference between CSMtLR and CSMGtLR as well
as between CSMtR and CSMGtR.\\

Note that in the gluon-gluon process, as in the $t\bar t Z$ case,
CSMvH produces no effect as no 3-boson coupling is involved.\\

With our choices of form factors described in Sect. 3
one gets the following results for the ordering, the details being shown
in Fig.7-9:\\

For $W^-_L$  one gets CSMvH/CSMtLR/CSMvt/CSMtR/CSMGtR/CSMGtLR in gluon-gluon and\\
CSMvH/CSMvt/CSMGtR/CSMtR/CSMtLR/CSMGtLR in $\gamma\gamma$,\\

for $W^-_T$, CSMvH/CSMtR/CSMvt/CSMtLR in gluon-gluon and\\
CSMvH/CSMtR/CSMtLR/CSMvt in $\gamma\gamma$,\\

and in the unpolarized case, CSMvH/CSMtLR/CSMtR/CSMvt/CSMGtR/CSMGtLR in gluon-gluon and
CSMvH/CSMtvt/CSMtR/CSMGtR/CSMGtLR/CSMtLR in $\gamma\gamma$.\\

\section{CONCLUSIONS}

In this paper we have shown how the processes of $t\bar t H, t\bar t Z, t\bar b W^-$
production in gluon-gluon and photon-photon collisions may indicate what type
of Higgs and top quark compositeness would be at the origin of a possible departure
with respect to SM predictions.\\
We have concentrated our analysis on the shapes
of the energy dependence of the cross sections possibly modified by specific
Higgs and top quark form factors satisfying or violating the CSM constraints.\\
Peculiar modifications appear in each process and especially
in the case of $Z_L$ and $W_L$ production due to the perturbations of
the specific gauge cancellations generated by such form factors.\\
In this study as well as in the previous ones for other processes
our purpose was to show what are the opportunities of testing CSM constraints.
In order to go further a specific CSM model should be available in
order to make precise predictions for all details of the corresponding
cross sections. They could then be compared to expected possibilities
at LHC and future colliders, see \cite{Richard}
and \cite{gammagamma}.\\

\newpage

\begin{figure}[p]

\[
\epsfig{file=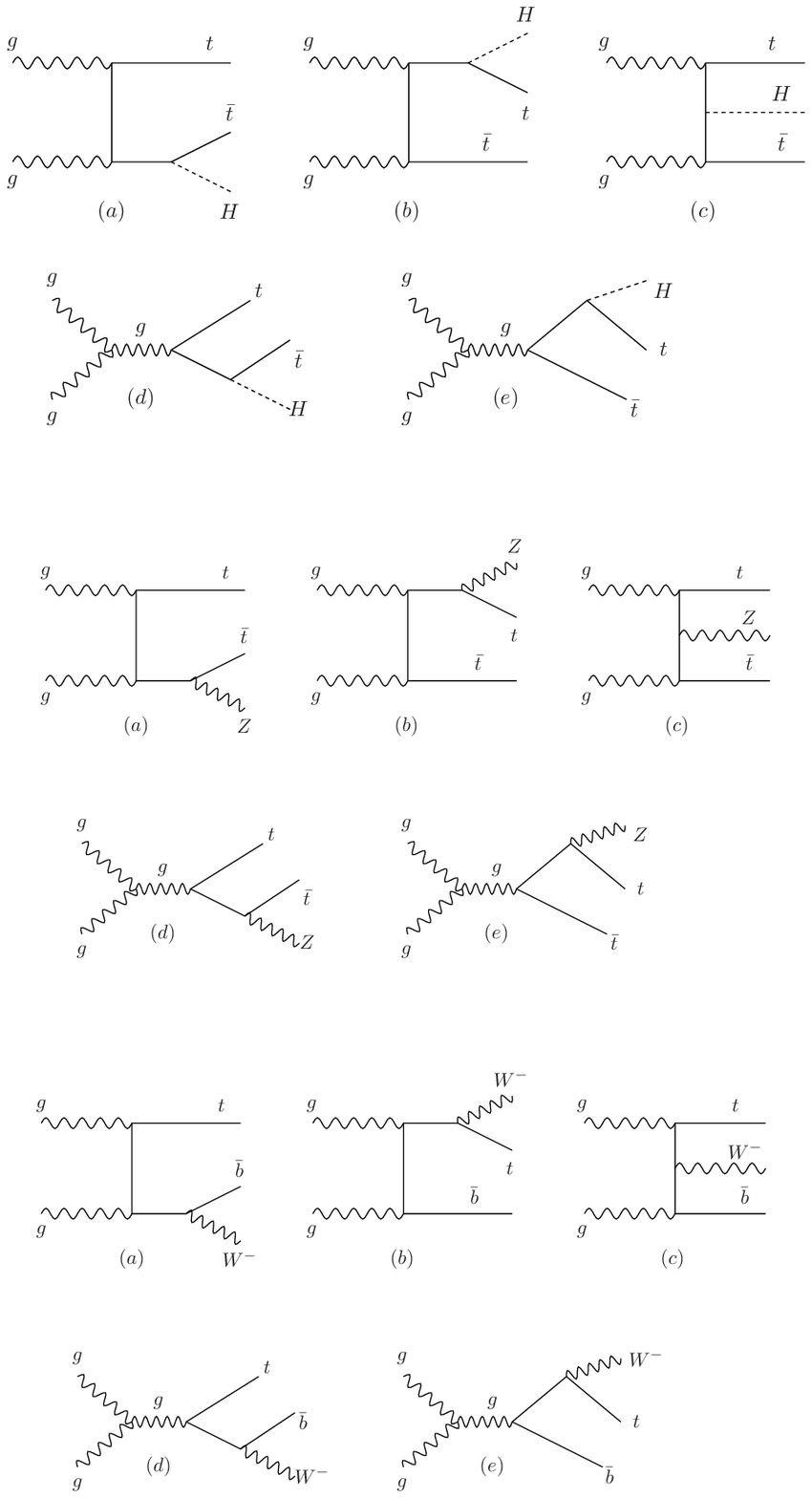, height=20.cm}
\]\\

\vspace{-1cm}
\caption[1] {Diagrams for gluon-gluon collisions}
\end{figure}

\clearpage

\begin{figure}[p]
\vspace{-3cm}
\[
\epsfig{file=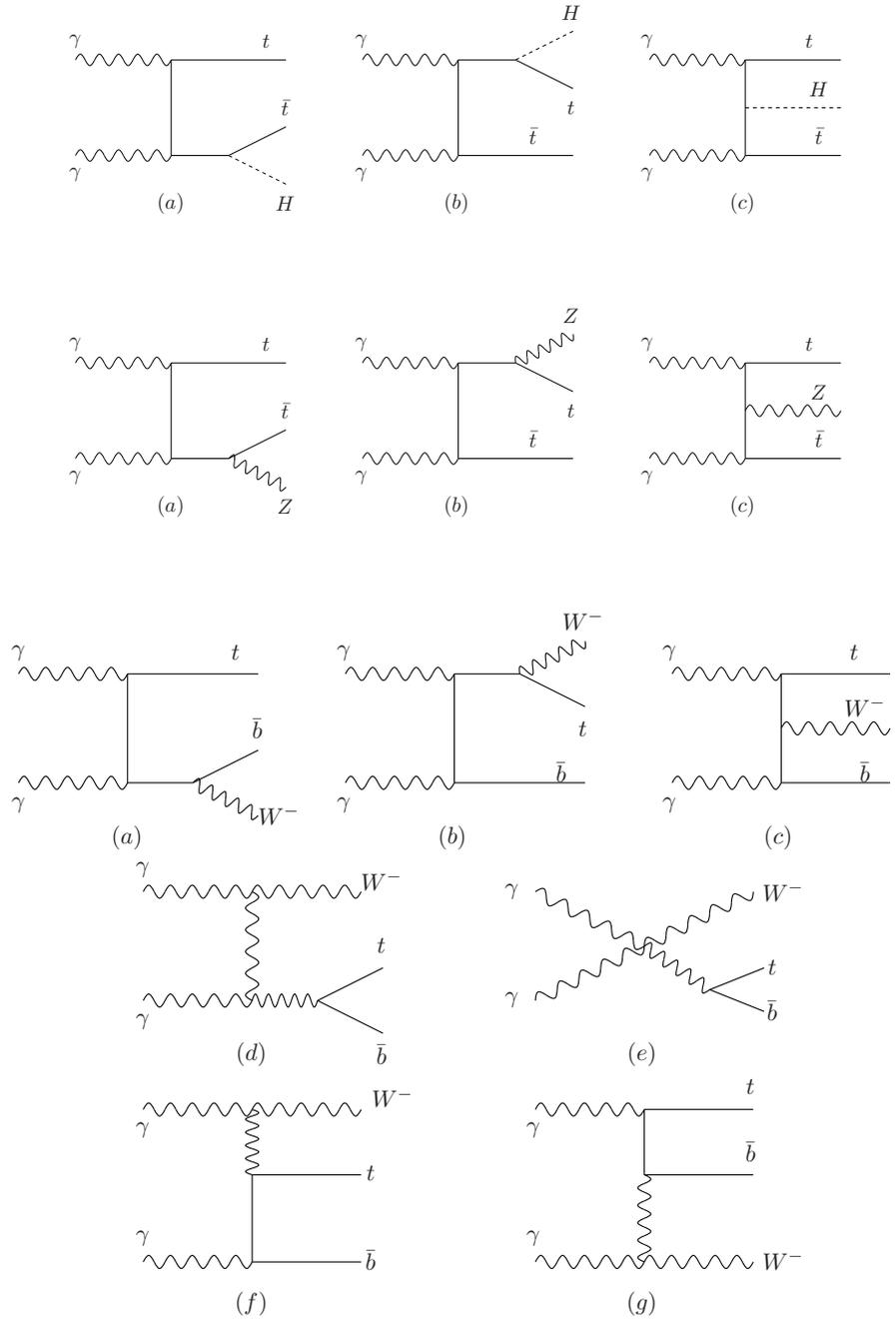, height=28.cm}
\]\\
\vspace{-6cm}
\caption[1] {\hspace{1cm} Diagrams for photon-photon collisions}
\end{figure}

\clearpage

\begin{figure}[p]
\[
\epsfig{file=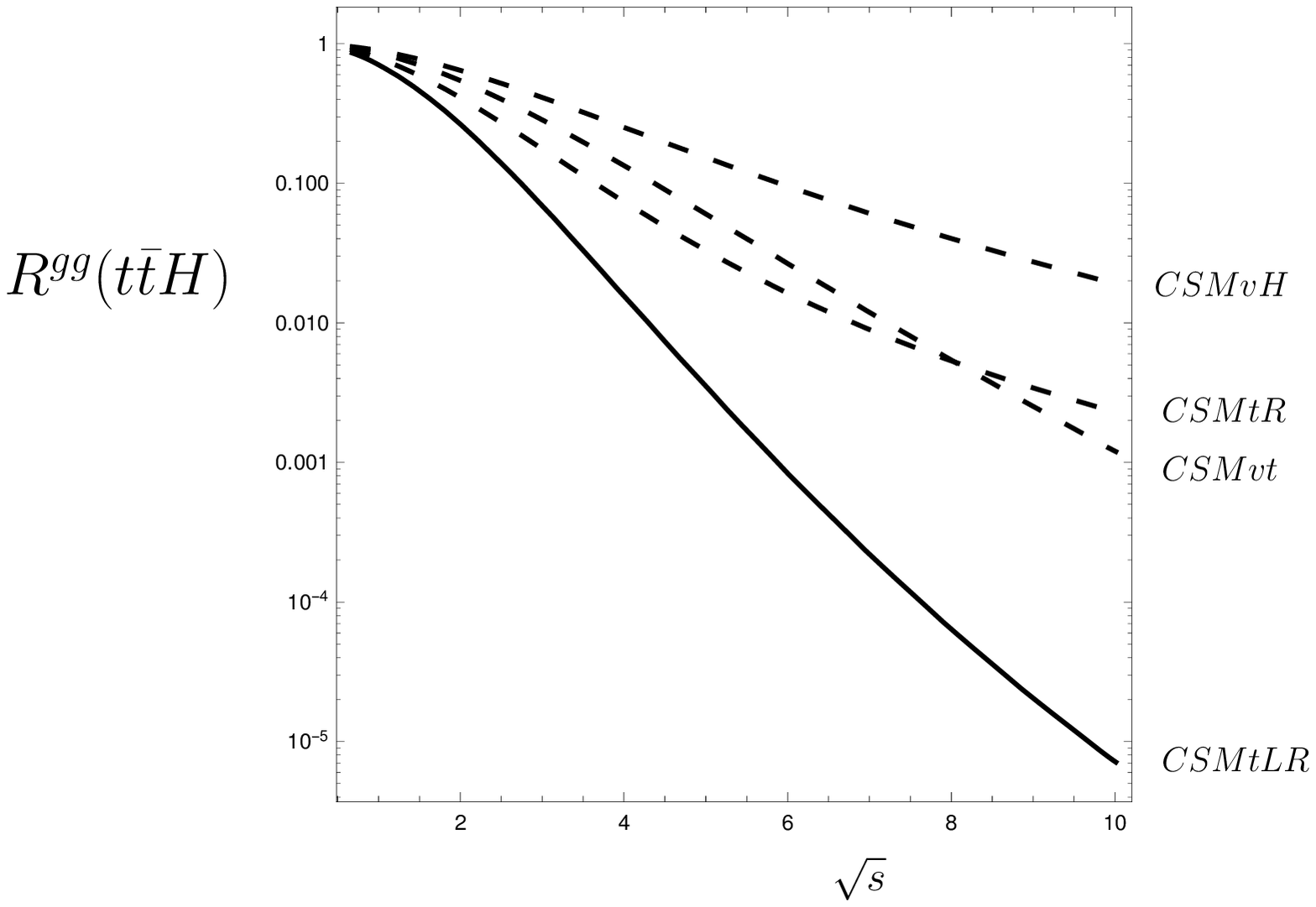, height=6.cm}
\epsfig{file=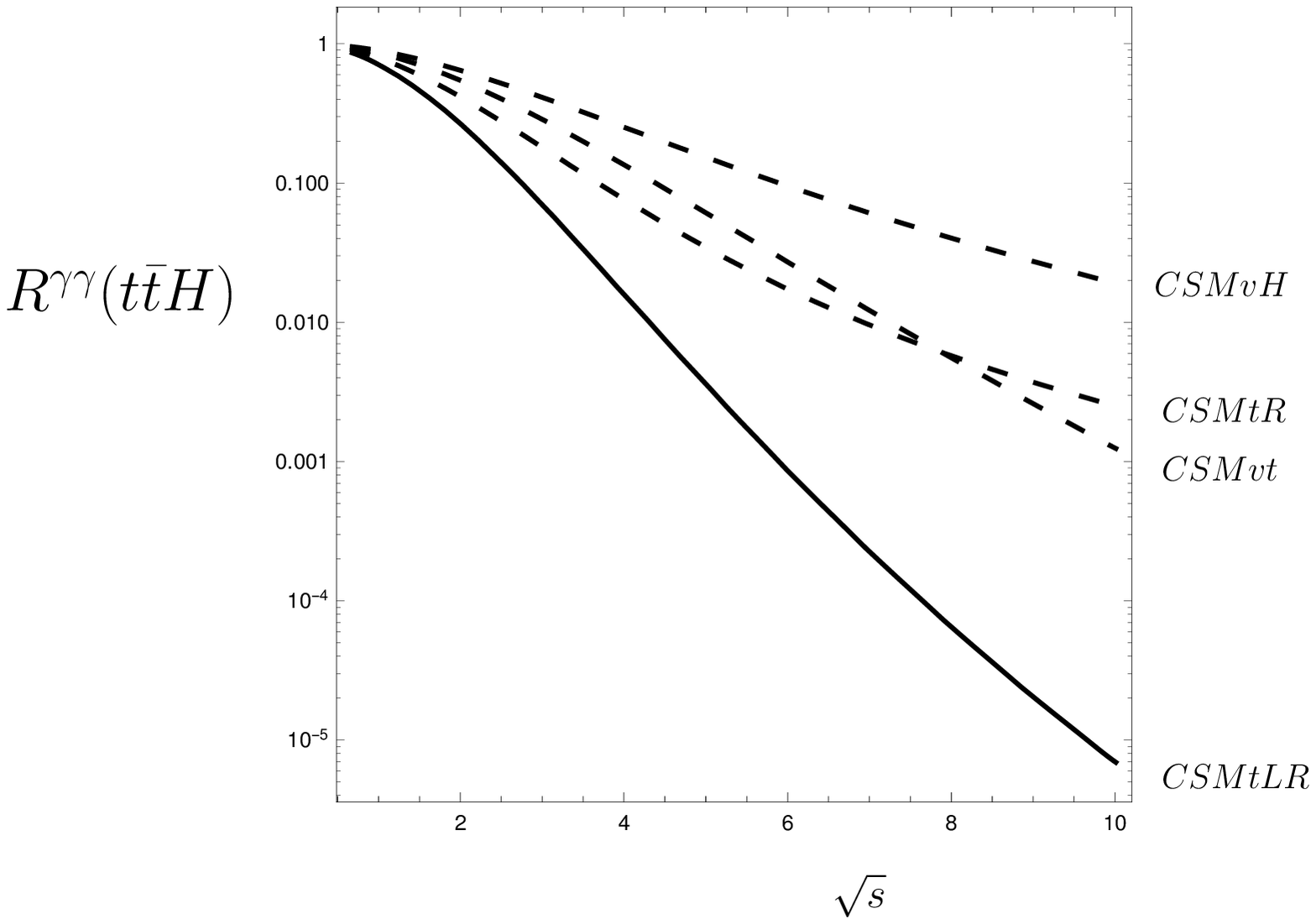, height=6.cm}
\]
\hspace{4.5cm}(a)\hspace{7cm}(b)
\vspace{1.5cm}

\caption[1] {Ratios for $t\bar t H$ production,
(a) in gluon-gluon collisions,
(b) in photon-photon collisions.}
\end{figure}

\clearpage

\begin{figure}[p]

\vspace{-1.5cm}
\[
\epsfig{file=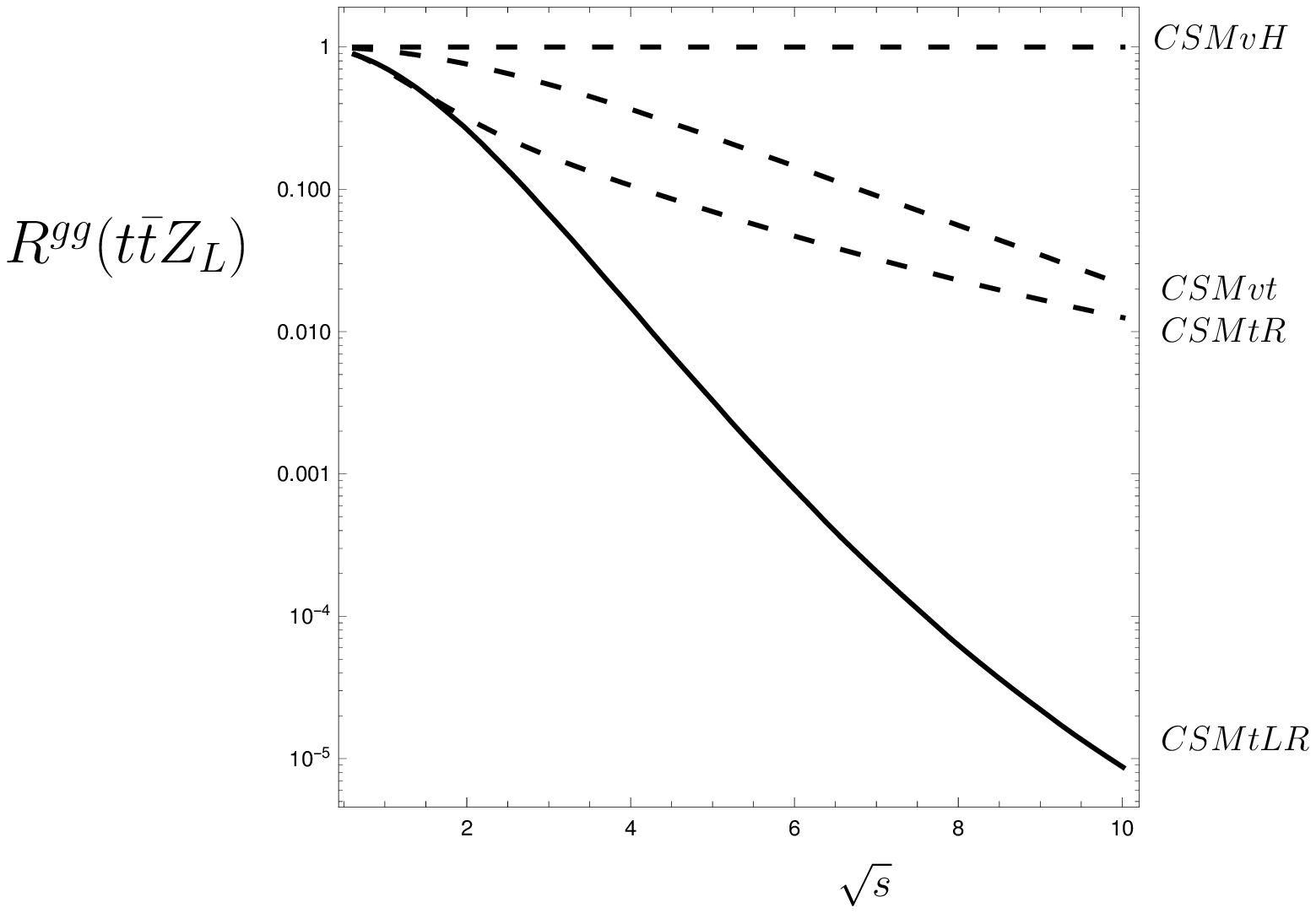, height=6.cm}
\epsfig{file=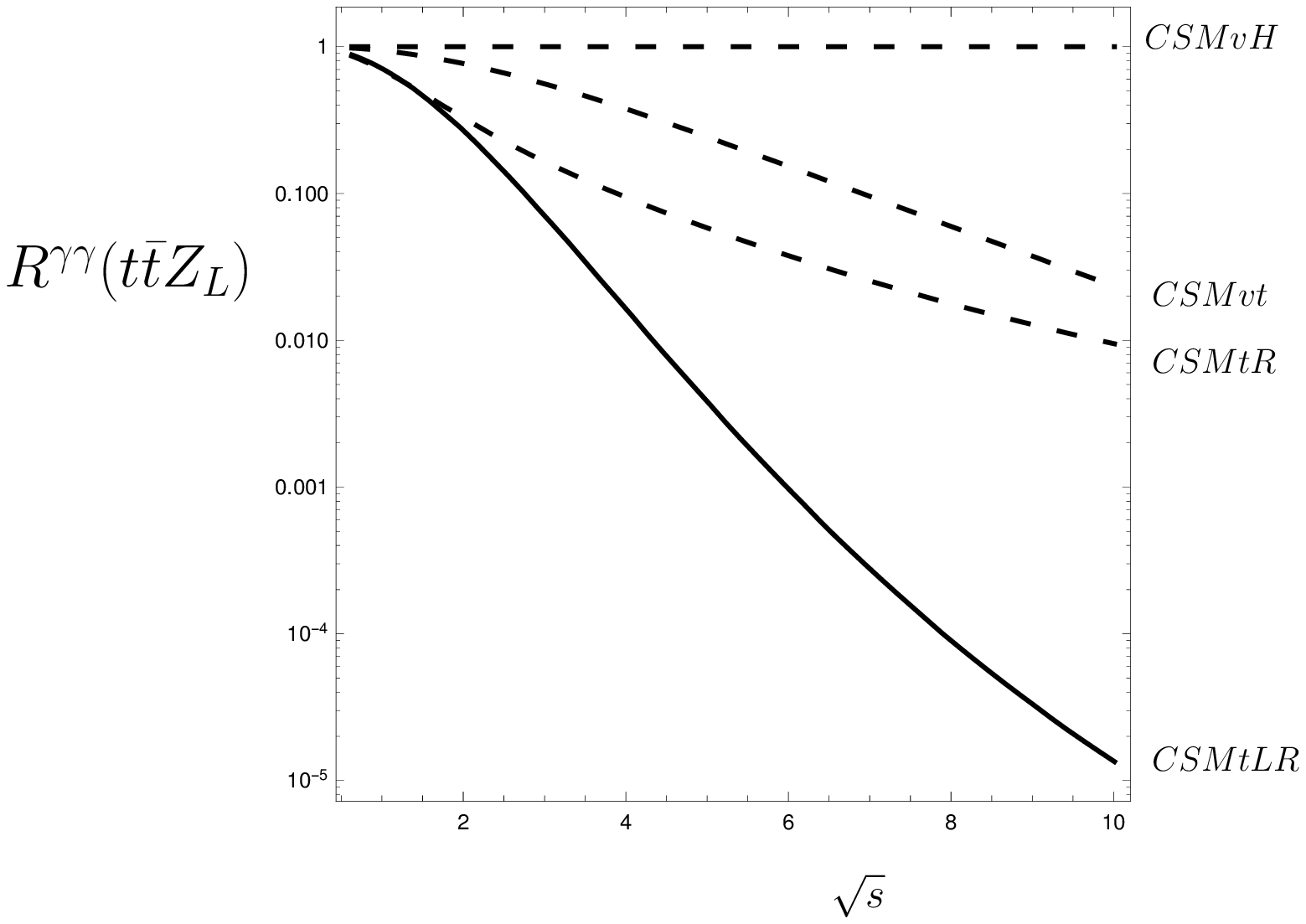, height=6.cm}
\]

\hspace{4.5cm}(a)\hspace{7cm}(b)
\vspace{1.5cm}

\caption[1] {Ratios for $t\bar t Z_L$ production,
(a) in gluon-gluon collisions,
(b) in photon-photon collisions.}
\end{figure}

\clearpage

\begin{figure}[p]
\[
\epsfig{file=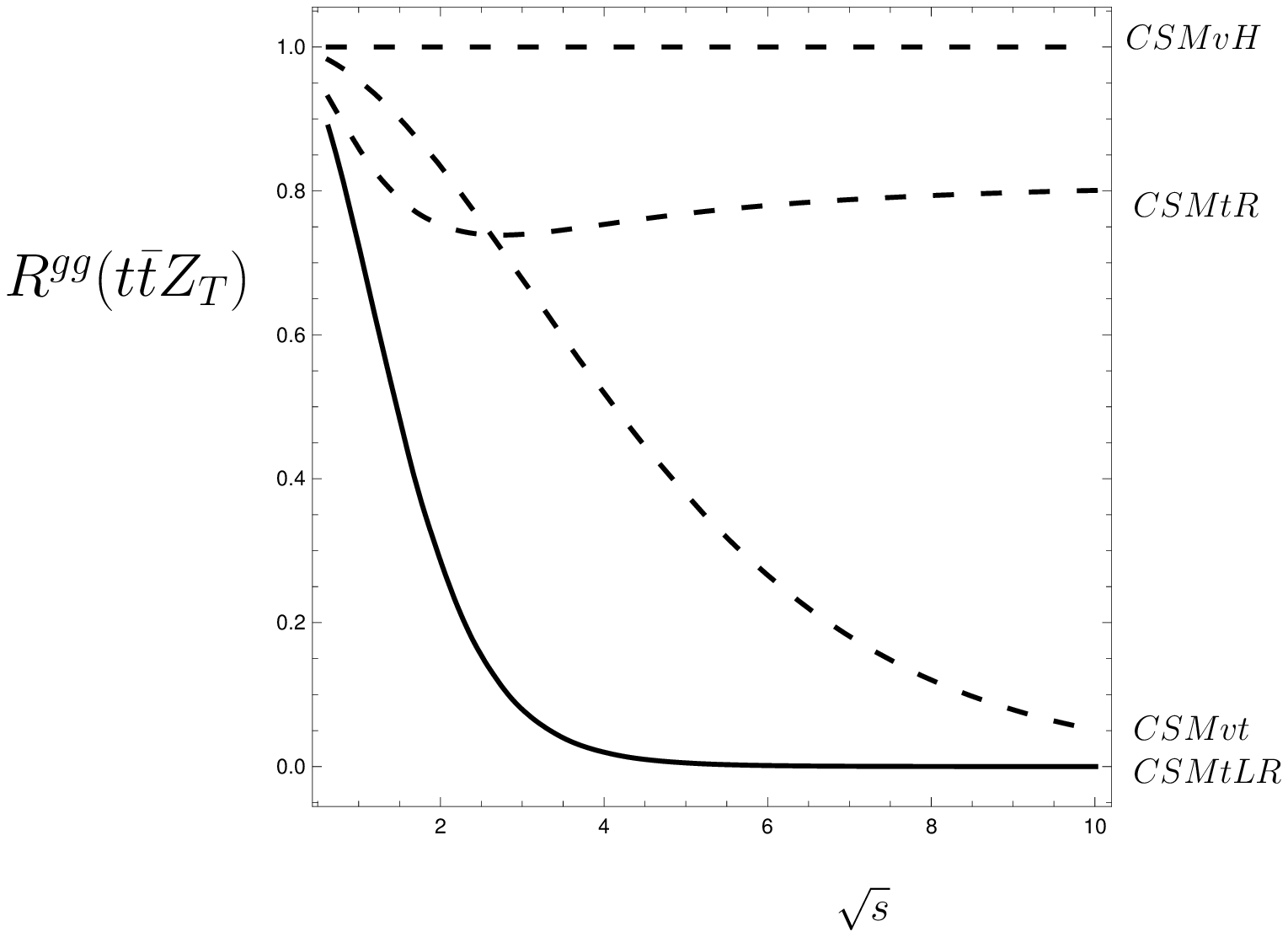, height=6.cm}
\epsfig{file=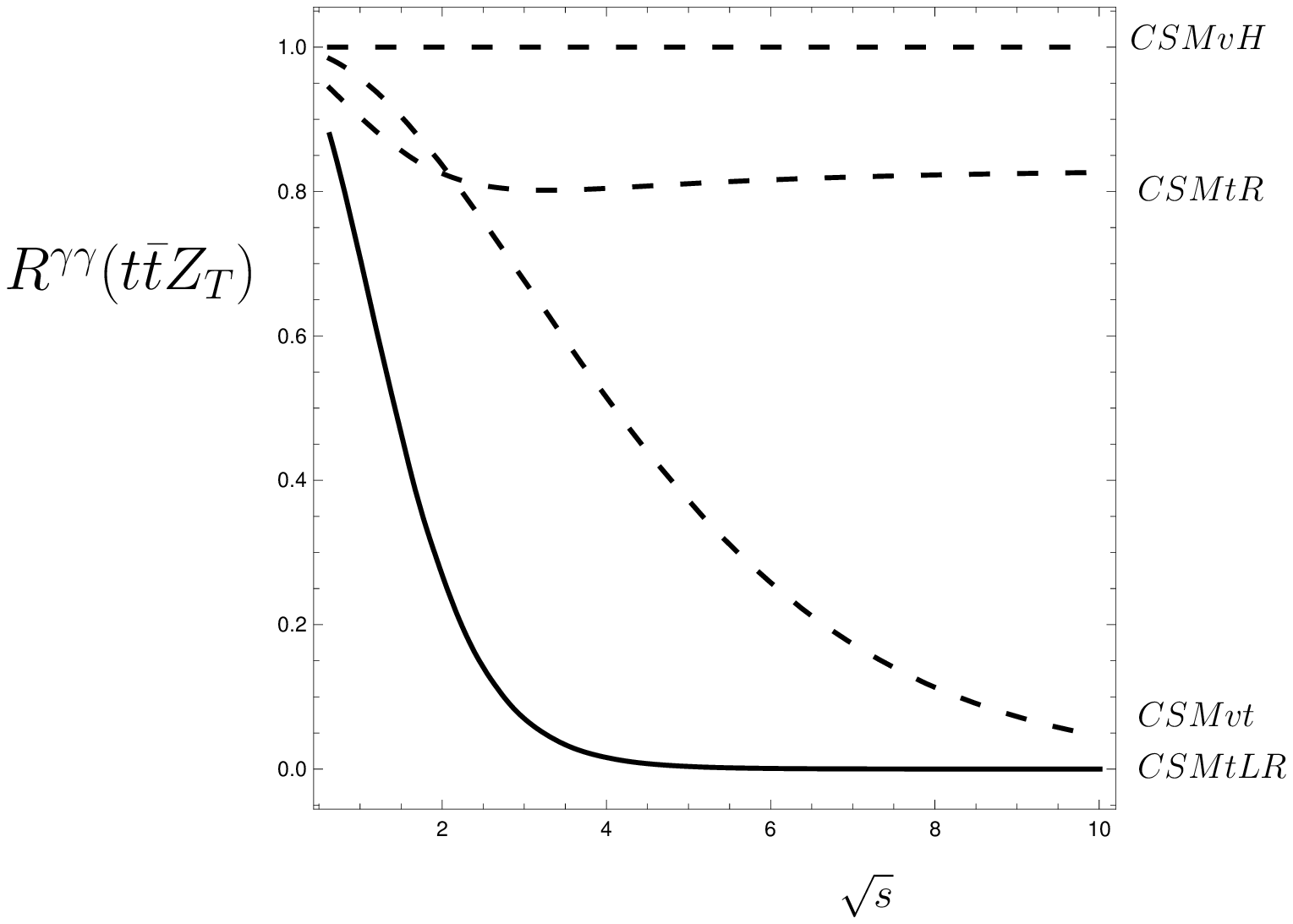, height=6.cm}
\]\\

\hspace{4.5cm}(a)\hspace{7cm}(b)
\vspace{1.5cm}

\caption[1] {Ratios for $t\bar t Z_T$ production,
(a) in gluon-gluon collisions,
(b) in photon-photon collisions.}
\end{figure}

\clearpage

\begin{figure}[p]
\[
\epsfig{file=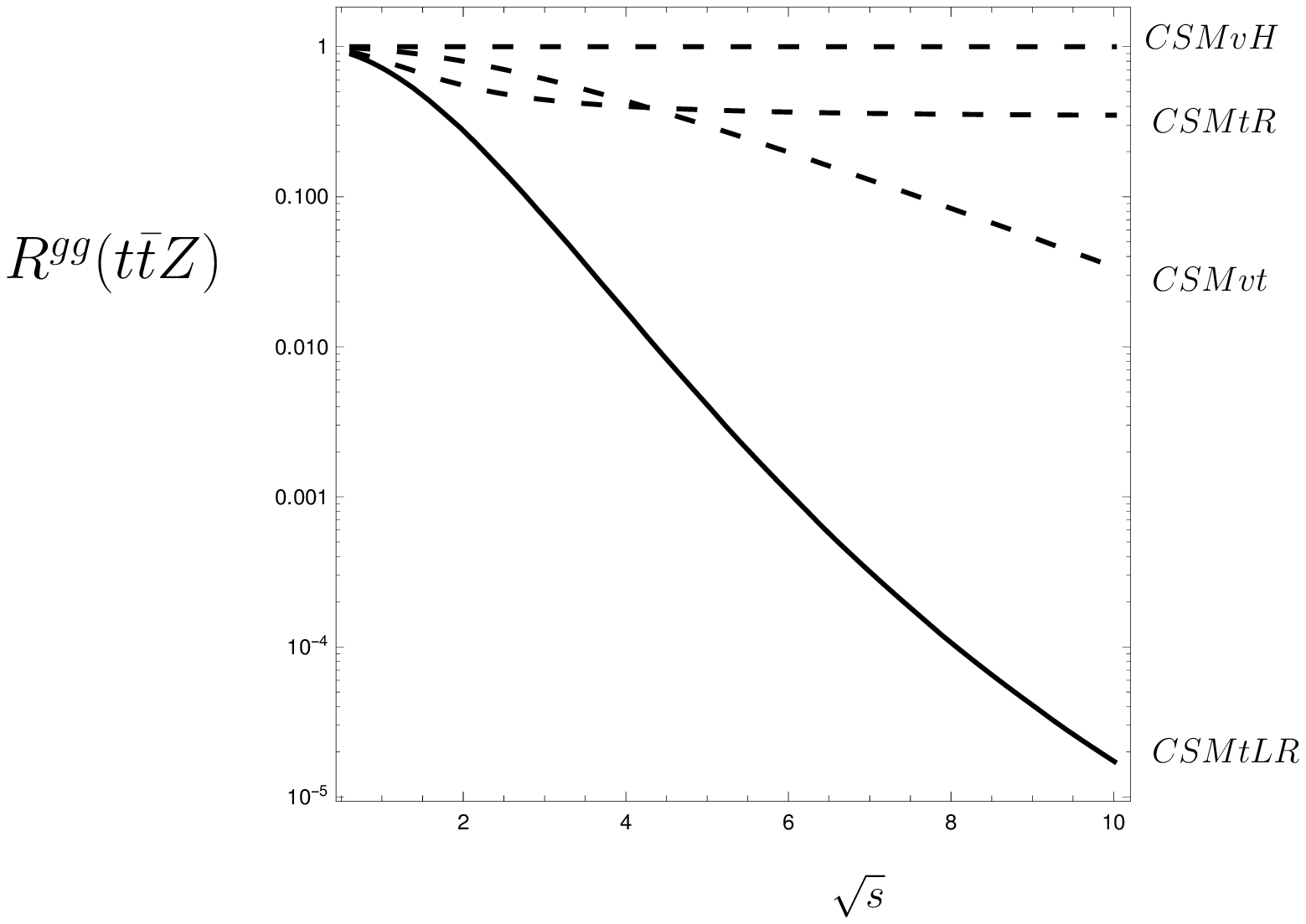, height=6.cm}
\epsfig{file=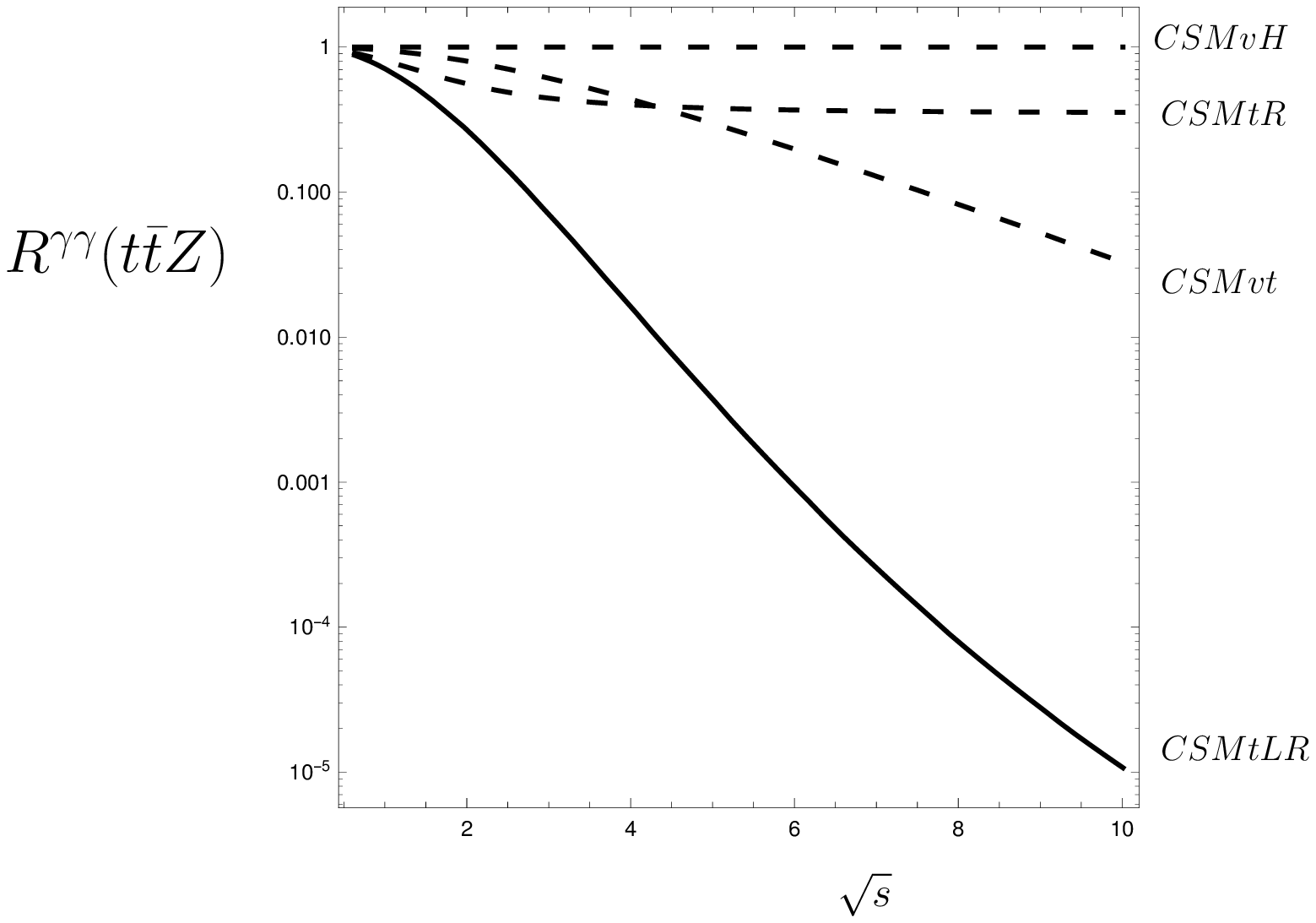, height=6.cm}
\]\\

\hspace{4.5cm}(a)\hspace{7cm}(b)
\vspace{1.5cm}

\caption[1] {Ratios for unpolarized $t\bar t Z$ production,
(a) in gluon-gluon collisions,
(b) in photon-photon collisions..}

\end{figure}
\clearpage

\begin{figure}[p]
\[
\epsfig{file=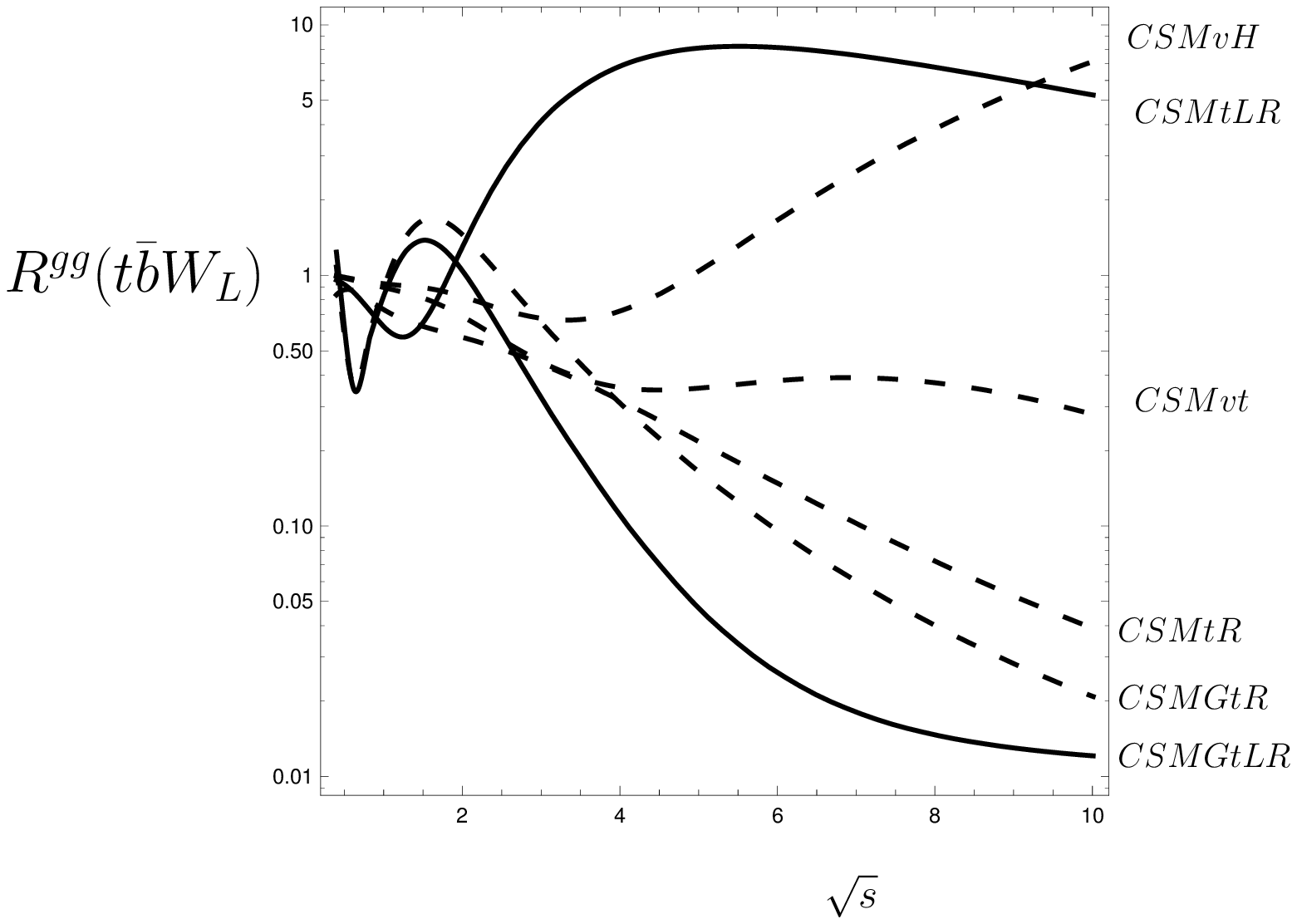, height=6.cm}
\epsfig{file=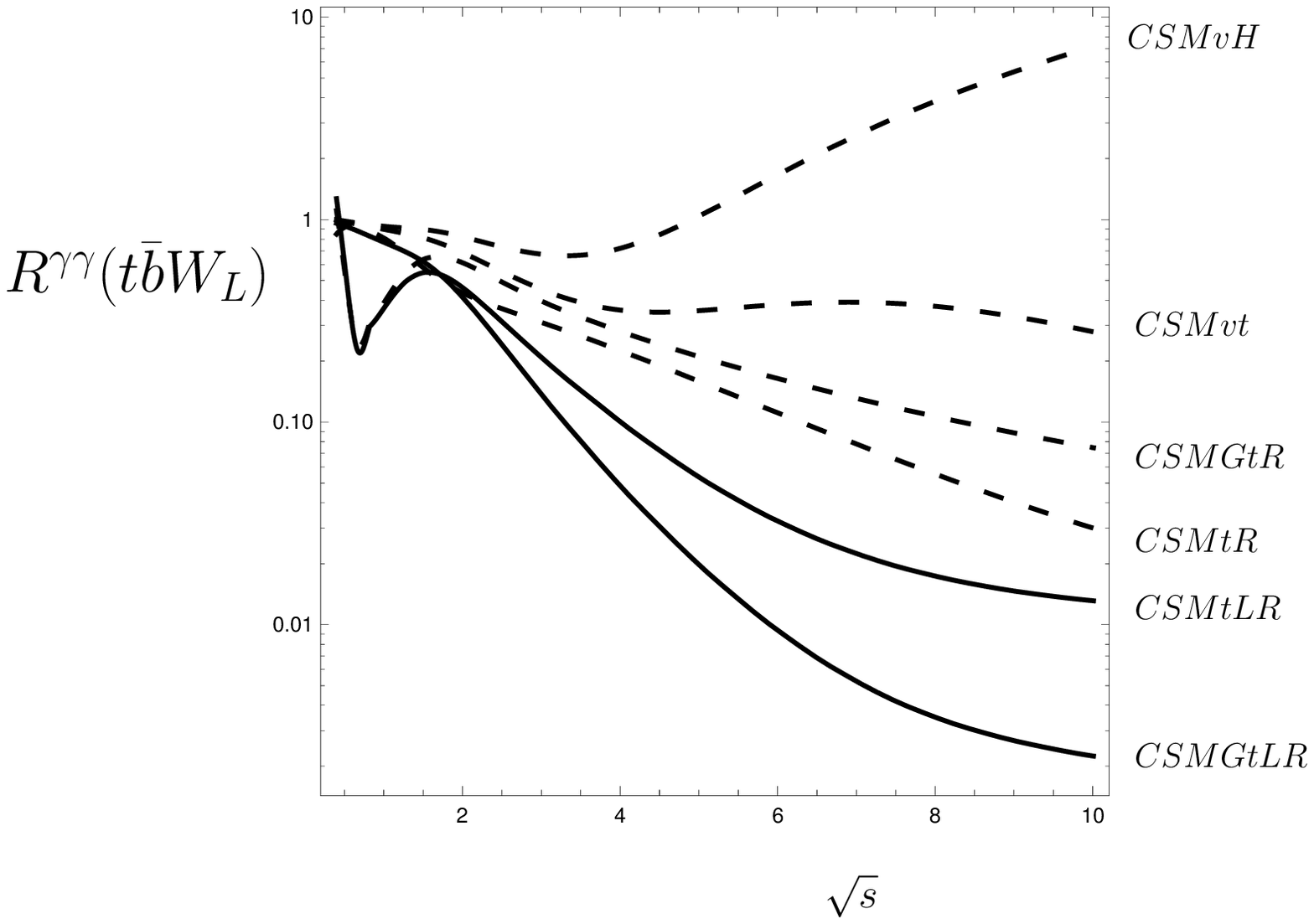, height=6.cm}
\]\\

\hspace{4.5cm}(a)\hspace{7cm}(b)
\vspace{1.5cm}

\caption[1] {Ratios for $t\bar b W^-_L$ production,
(a) in gluon-gluon collisions,
(b) in photon-photon collisions.}
\end{figure}
\clearpage

\begin{figure}[p]
\[
\epsfig{file=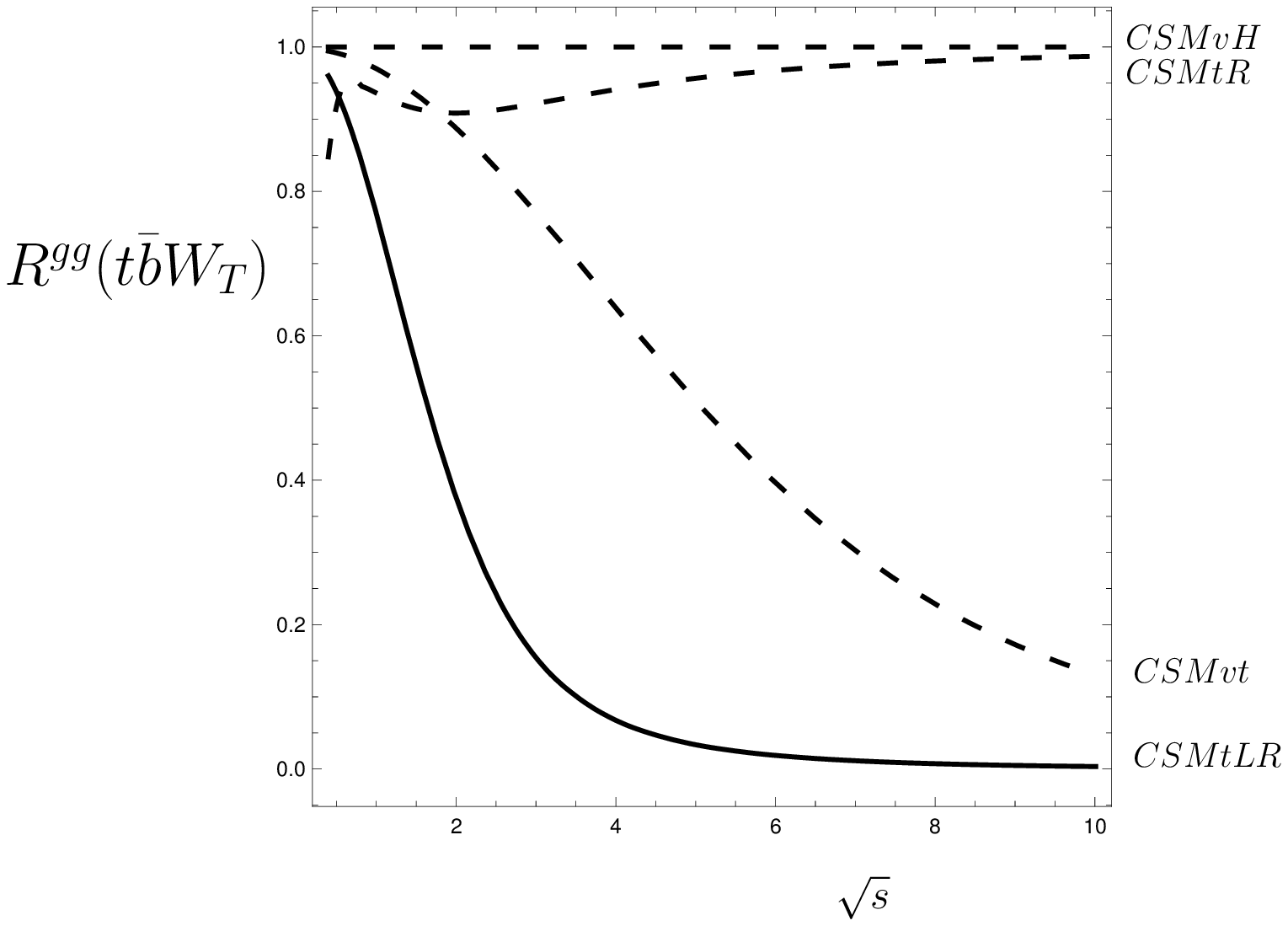, height=6.cm}
\epsfig{file=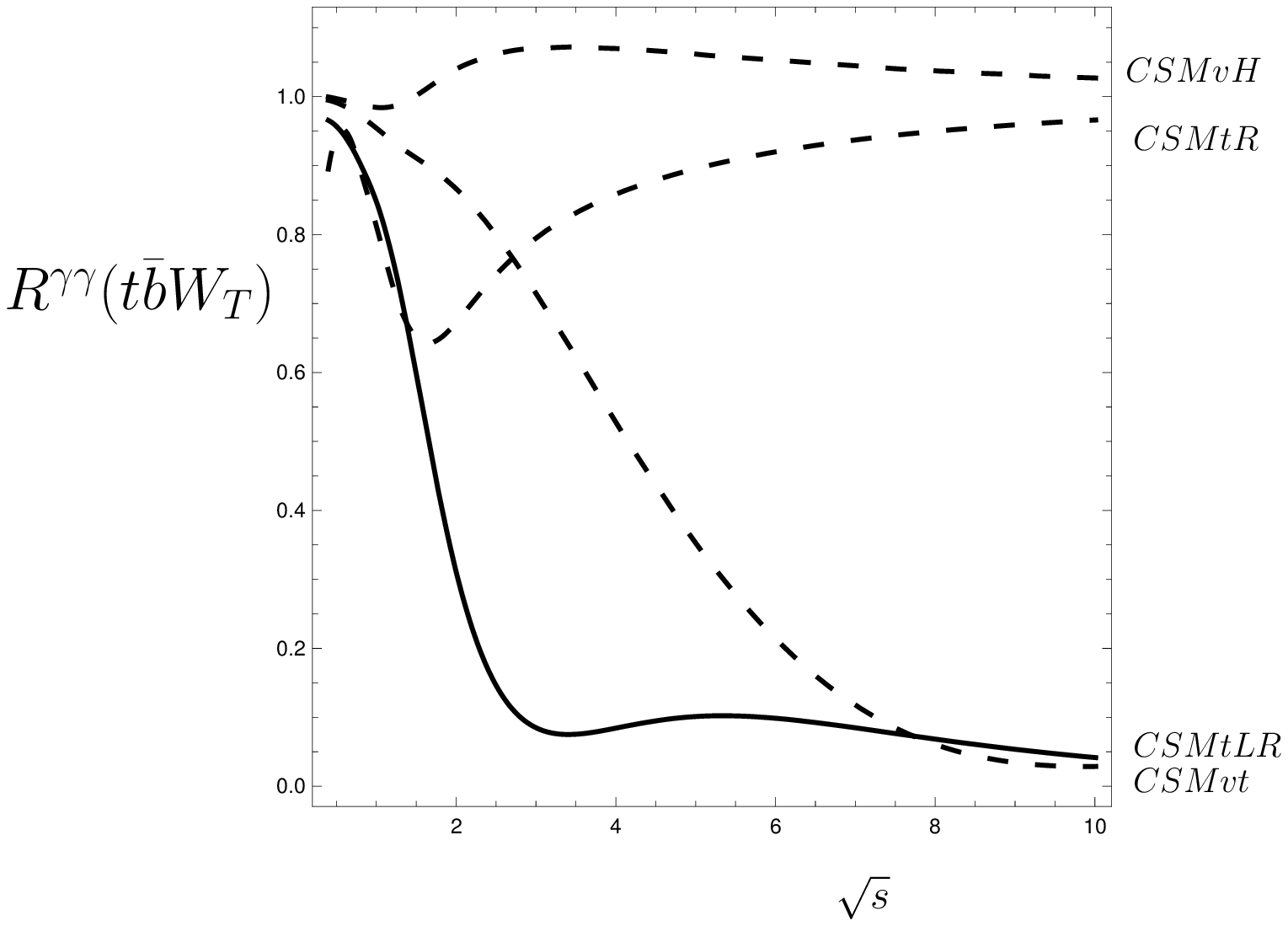, height=6.cm}
\]\\

\hspace{4.5cm}(a)\hspace{7cm}(b)
\vspace{1.5cm}

\caption[1] {Ratios for $t\bar b W^-_T$ production,
(a) in gluon-gluon collisions,
(b) in photon-photon collisions.}

\end{figure}

\begin{figure}[p]
\[
\epsfig{file=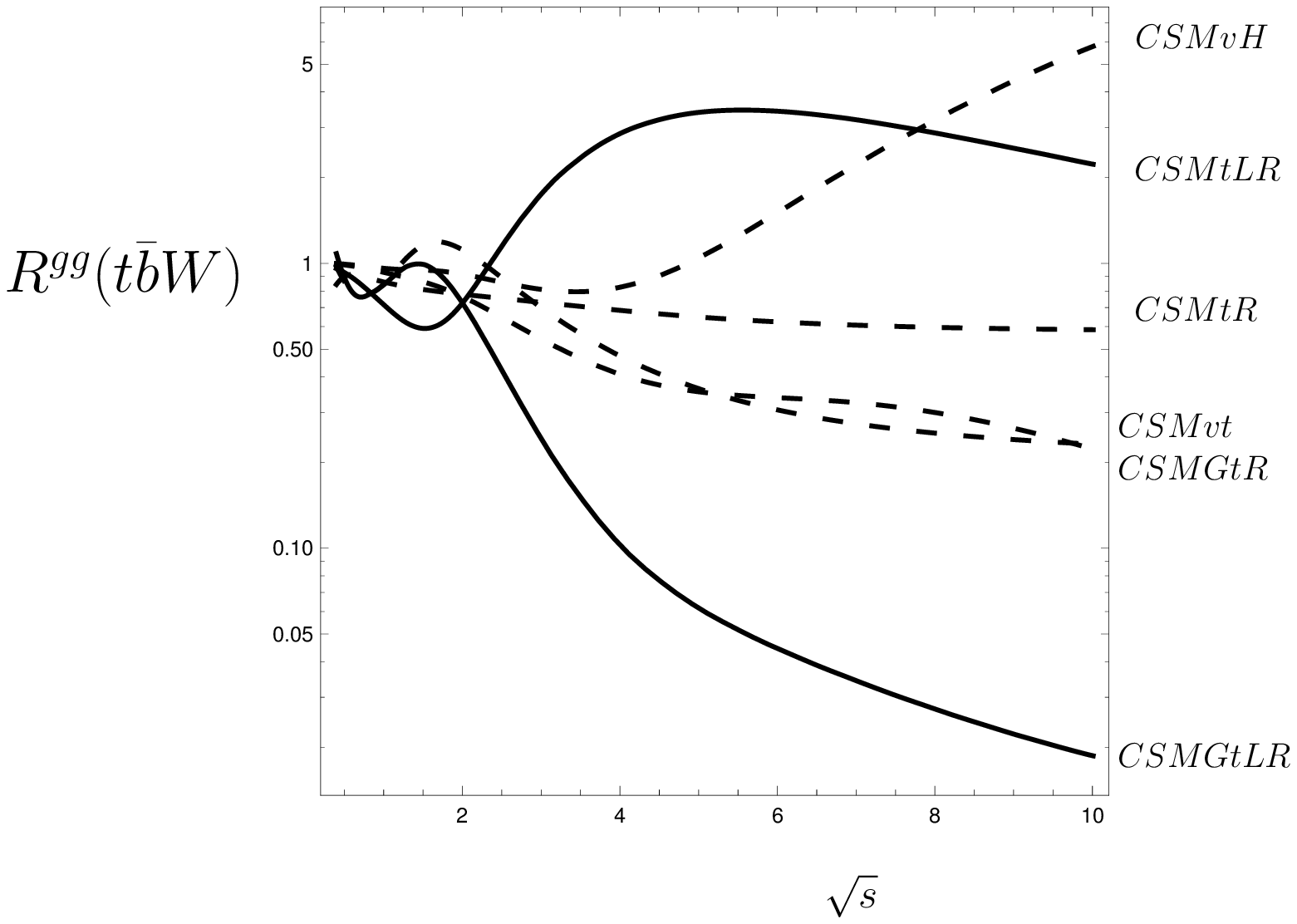, height=6.cm}
\epsfig{file=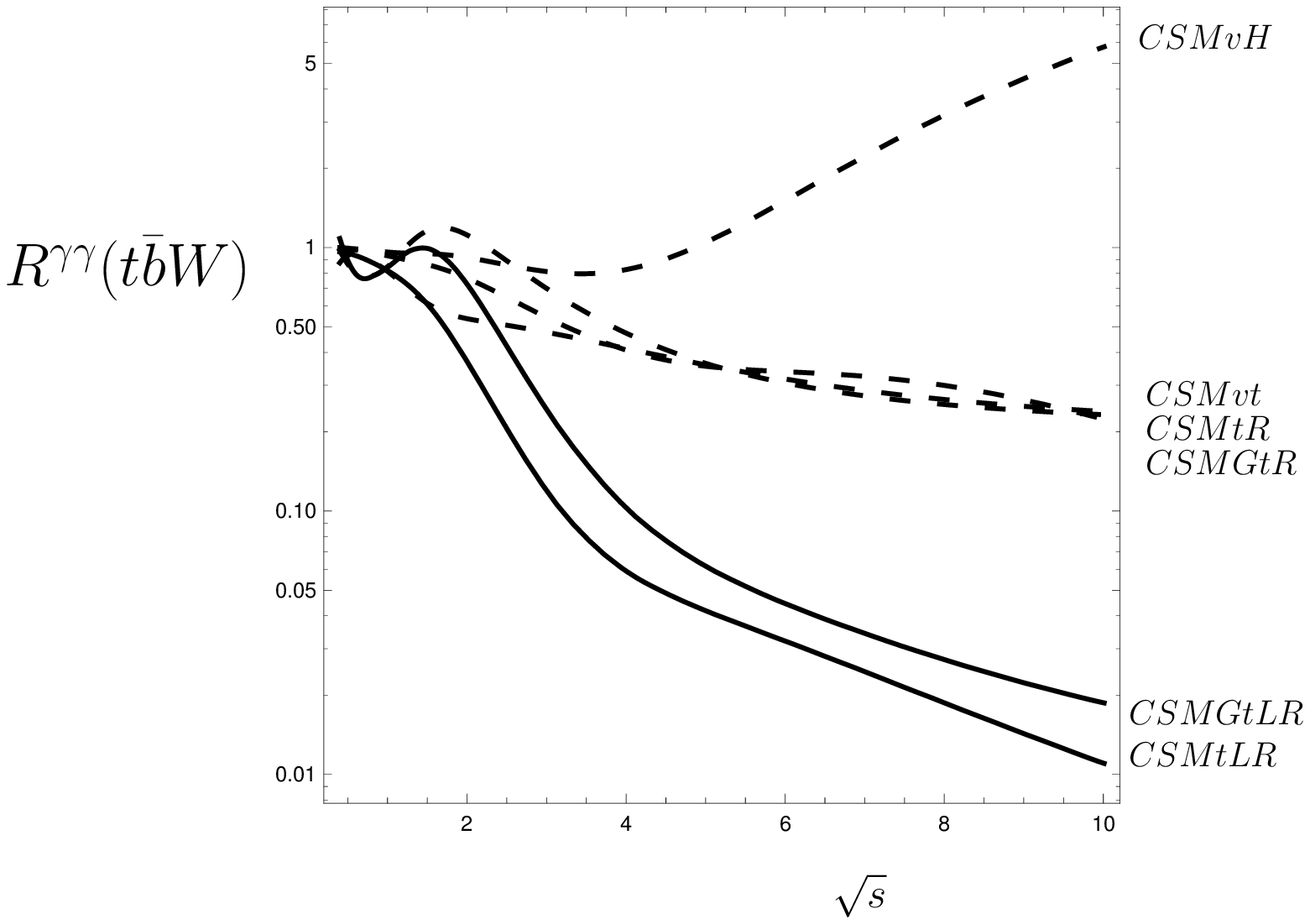, height=6.cm}
\]\\

\hspace{4.5cm}(a)\hspace{7cm}(b)
\vspace{1.5cm}

\caption[1] {Ratios for unpolarized $t\bar b W^-$ production,
(a) in gluon-gluon collisions,
(b) in photon-photon collisions.}
\end{figure}

\end{document}